# Optic phonons and the soft mode in 2H-NbSe$_2$

F. Weber[1], R. Hott[1], R. Heid[1], K.-P. Bohnen[1], S. Rosenkranz[2], J.-P. Castellan[2], R. Osborn[2], A. H. Said[3], B. M. Leu[3] and D. Reznik[4]

[1] Institute of Solid State Physics, Karlsruhe Institute of Technology, 76021 Karlsruhe, Germany
[2] Materials Science Division, Argonne National Laboratory, Argonne, Illinois, 60439, USA
[3] Advanced Photon Source, Argonne National Laboratory, Argonne, Illinois, 60439, USA
[4] Department of Physics, University of Colorado at Boulder, Boulder, Colorado, 80309, USA

**We present an investigation of the lattice dynamics of the charge-density-wave compound 2H-NbSe$_2$. We analyze the precise nature of the wave vector dependent electron-phonon coupling (EPC) and derive the bare dispersion of the charge-density-wave (CDW) soft phonon mode using inelastic x-ray scattering combined with ab-initio calculations. Experimentally, phonon modes along the Γ – M line, i.e. q = (h, 0, 0) with 0 ≤ h ≤ 0.5, with the same longitudinal symmetry (Σ$_1$) as the CDW soft mode were investigated up to 32 meV. In agreement with our calculations we observe significant EPC in the optic modes at h ≤ 0.2. We analyze the EPC in the optic as well as acoustic modes and show that the q dependences stem from scattering processes between two bands at the Fermi surface both having Nb 4d character. Finally, we demonstrate that the soft mode dispersion at T = 33 K (= T$_{CDW}$) can be well described on the basis of a strongly q dependent EPC matrix element and an acoustic-like bare phonon dispersion in agreement with observations near room temperature.**

## I. Introduction

Charge-density-wave (CDW) formation is one of the most common phenomena in solid state physics and relevant to a number of important issues in condensed matter physics, such as the role of stripes in cuprate superconductivity [1] and charge fluctuations in the colossal magnetoresistive manganites [2]. Static CDW order, i.e. a periodic modulation of the electronic density, can only be stabilized in case of a non-zero electron-phonon coupling, specifically, coupling of phonons to electrons in the conduction bands. Hence, the electronic modulation is accompanied by a lattice distortion involving a soft phonon mode with a zero energy at **q** = **q**$_{CDW}$ and T = T$_{CDW}$.

2H-NbSe$_2$ is a prototypical CDW compound. It was originally investigated over four decades ago as one of the first layered materials, in which superconductivity was observed (T$_{SC}$ (2H-NbSe$_2$) = 7.2 K) [3]. Only afterwards it was realized that 2H-NbSe$_2$ undergoes a CDW phase transition already at T$_{CDW}$ = 33 K [4], however, the exact distortion pattern at T < T$_{CDW}$ is still a subject of current research [5]. Original ideas on the origin of the CDW formation centered on the Fermi surface nesting, however, subsequent experiments found that CDW in some compounds appears without strong FS nesting [6]. Early on, an alternative mechanism based on a q-dependent enhancement of the electron-phonon coupling matrix element, $g_q$, has been proposed. A prominent role of EPC and, in particular, the wave vector dependence of $g_q$ have been suggested [7,8,9]. Experimentally, however, the small size of 2H-NbSe$_2$ single crystals allowed only a limited investigation of the CDW soft phonon mode close to T$_{CDW}$ by inelastic neutron scattering [10-12]. Earlier, we reported high-resolution inelastic x-ray scattering experiments showing evidence that in 2H-NbSe$_2$ the wave vector dependence of $g_q$ is indeed at the origin of the CDW transition [13].

Any realistic model of soft phonons in CDW compounds must begin with understanding the bare phonon dispersion, $\omega_{bare}(q)$, which is the dispersion without the interaction of the phonon with electrons in the conduction bands. It takes into account the screening of the ionic movements by the strongly bound core electrons but not the more subtle effects due to electronic scattering processes near the Fermi surface. Extracting $\omega_{bare}(q)$ is not a trivial task, because it cannot be measured directly. Furthermore, it is necessary to go beyond the simple assumption that the soft phonon must derive from an acoustic branch, because optic phonon can also soften to zero energy if they are coupled to conduction electrons strongly enough.

In this report we derive the bare phonon dispersion $\omega_{bare}(q)$ of the CDW soft mode in NbSe$_2$ from a detailed analysis of the wave vector dependent EPC, the correlated electronic scattering processes and the phonon displacement patterns. Since the soft phonon mode is not necessarily acoustic-like, optic phonons of the same symmetry as the soft mode had to be investigated as well. We have measured phonons with Σ$_1$ symmetry up to 32 meV providing evidence of EPC in the optical branches at small wave vectors. These results are in good agreement with our ab-initio calculations and we used the latter to further analyse the observed EPC in the optic as well as acoustic phonons. We found that the wave vector dependence of EPC in the investigated phonons is primarily due to electronic scattering processes between two Nb-4d derived bands at the Fermi surface. Further, the phonon patterns do not indicate an exchange of eigenvectors between optic and acoustic phonons. Accordingly, the soft mode at T = T$_{CDW}$ is acoustic-like.

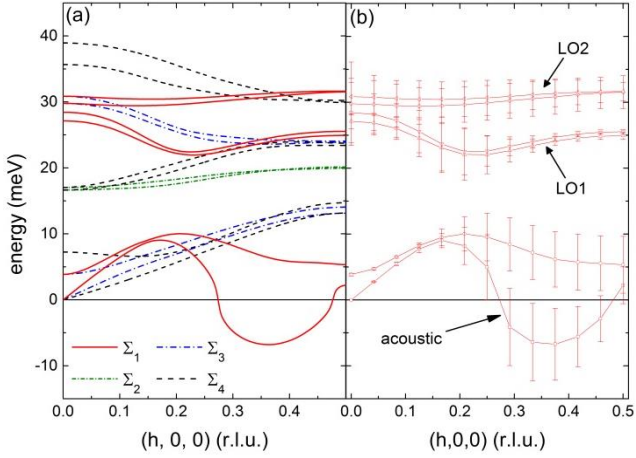

Figure 1: *(a)* Calculated phonon dispersion along the (100) direction in NbSe$_2$. Different line types (colors) denote branches with different symmetries (see legend). *(b)* Branches with longitudinal symmetry ($\Sigma_1$) (solid lines in *(a)*). Labels are given corresponding to the discussion. Vertical bars denote the calculated electronic contribution to the phonon line width $2\gamma$ scaled by a factor of 10 for the sake of visibility.

## II. Theory

Calculations using density-functional-perturbation-theory (DFPT) were performed in the framework of the mixed basis pseudopotential method.[14] The exchange-correlation functional was treated in the local-density approximation (LDA). Norm-conserving pseudopotentials for Nb and Se were constructed including 4s and 4p semicore states in the valence space in the case of Nb. The deep potentials can be efficiently treated in the mixed-basis scheme, which combines local functions together with plane waves for the representation of the valence states. Local functions of s, p, and d symmetry at the Nb sites and of s and p symmetry on the Se sites, respectively, were combined with plane waves up to 24 Ry.

Phonon energies and electron-phonon coupling were calculated using the linear response technique or density functional perturbation theory (DFPT)[15] in combination with the mixed-basis pseudopotential method.[16] To resolve fine features related to the Fermi surface geometry, Brillouin-zone (BZ) integrations were performed with a dense hexagonal 24 × 24 × 8 **k**-point mesh (244 points in the irreducible BZ). The standard smearing technique was employed with a Gaussian broadening of 0.1 eV. Tests with the denser **k**-point mesh confirmed sufficient convergence for both phonon energies and line widths. All results were obtained for the fully optimized hexagonal structure (a=b=3.40 Å, c=12.09 Å).

Imaginary phonon energies, shown as the negative roots of the square phonon energies in Fig. 1, appear in the calculation, because of an anharmonic double-well potential reflecting the CDW instability. At zero temperature the lattice will distort into a CDW and sit in one of the minima. However, the calculation assumes that the lattice is undistorted, i.e. that it sits in the middle (i.e. local maximum) of the double well potential. Negative curvature of the potential at atomic positions of the high temperature structure is what gives an imaginary calculated energy for the phonon whose eigenvector is close to the CDW distortion.

## III. Experiment

The IXS experiments were carried out at the XOR 30-ID HERIX beamline of the Advanced Photon Source, Argonne National Laboratory, with a focused beam size of 30 µm. The incident energy was 23.724 keV [17] and the horizontally scattered beam was analysed by a set of spherically curved silicon analysers (Reflection 12 12 12) [18]. The full width at half maximum (FWHM) of the energy and wave vector space resolution was about 1.5 meV and 0.066 Å$^{-1}$, respectively, where the former is experimentally determined by scanning the elastic line of a piece of plastic and the latter is calculated from the experiment geometry and incident energy. The components (Q$_h$, Q$_k$, Q$_l$) of the scattering vector are expressed in reciprocal lattice units (r.l.u.) (Q$_h$, Q$_k$, Q$_l$ = (h*2π/a, k*2π/a, l*2π/c) with the lattice constants a = b = 3.443 Å and c = 12.55 Å of the hexagonal unit cell, space group P6$_3$/mmc. Measurements were made in the constant-wave vector **Q** mode, i.e. as energy scans at constant wave vector **Q** = **τ** + **q**, where **τ** is a reciprocal lattice point and **q** the reduced wave vector. Measurements were done in the Brillouin zones adjacent to **τ** = (3, 0, 0) and (3, 0, 1), i.e. **Q** = (3-*h*, 0, 0) and (3-*h*, 0, 1). We used a high-quality single crystal sample of about 50 mg (2 x 2 x 0.05 mm³) with a T$_{CDW}$ of 33 K determined from the temperature dependence of the superlattice reflections [13] in agreement with previous results [19]. The sample was mounted in a closed cycle refrigerator and measurements reported here were done at various temperatures 33 K ≤ T ≤ 250 K.

Measured energy spectra were fitted using a pseudo-Voigt function for the elastic line with a variable amplitude and fixed line shape established by scanning through the CDW superlattice peak at T = 8 K and reference scans of a piece of plastic. Phonon peaks were fitted by a damped harmonic oscillator (DHO) function [20]

$$S(Q,\omega) = \frac{[n(\omega)+1]Z(Q)4\omega\Gamma/\pi}{[\omega^2 - \widetilde{\omega}_q^2]^2 + 4\omega^2\Gamma^2} \quad (1)$$

where Q and ω are the wave vector and energy transfer, respectively, n(ω) is the Bose function, $\Gamma$ is the imaginary part of the phonon self-energy, $\widetilde{\omega}_q$ is the phonon energy renormalized by the real part of the phonon self-energy and Z(Q) is the phonon structure factor. This function covers the energy loss and energy gain scattering by a single line shape and was convoluted with the experimental resolution. The intensity ratio of the phonon peaks at E = ±$\omega_q$ is fixed by the principle of detailed balance. The energy $\omega_q$, of the damped phonons is obtained from the fit parameters of the DHO function by $\omega_q = \sqrt{\widetilde{\omega}_q^2 - \Gamma^2}$ (Ref. [21]). Here, $\widetilde{\omega}_q$ is the phonon energy renormalized by the real part of the susceptibility, $Re\ \chi$, whereas $\omega_q$ is renormalized by both the real and imaginary part of the susceptibility.

## IV. Results
### IV.1 Density functional perturbation theory

Our calculated phonon dispersions are in good agreement with previous calculations [22]. In Fig. 1(a) we show the dispersion along the crystallographic (100) direction including the CDW wave vector $\mathbf{q}_{CDW}$ = (0.329,0,0) [19]. Due to the double layered structure within the hexagonal unit cell of NbSe$_2$, we observe pairs of branches with very small differences in absolute energy and dispersion except for the acoustic modes. As reported previously [13], our calculations predict a broad range of wave vectors with imaginary phonon energies indicating the structural instability. Nonetheless we discuss our results within the undistorted high temperature structure because we are interested in the lattice dynamics leading to the phase transition at and just above the phase transition.

Fig. 1(b) focuses on the longitudinal branches along the (100) direction having $\Sigma_1$ symmetry. Here, we also plot the momentum and energy resolved calculated electronic contribution to the line width of the phonons, which is a direct measure of the EPC. Sizeable contributions of EPC to the phonon line widths for the longitudinal acoustic (LA) and optic (LO) phonon branches are calculated. In the two highest LO modes we see very little wave vector dependence of EPC, whereas the two LO branches starting at the zone centre at 27.13 meV and 28.43 meV exhibit a clear decrease of EPC along the $\Gamma - M$ line (Fig. 1b). Moreover, the EPC of the LA and lowest LO branches strongly increase in the wave vector range where the line widths of the LO branches starting out at 27.13 meV and 28.43 meV are reduced (h = 0.2 – 0.25 r.l.u.). We point out that the wave vectors where the dispersion of the LA mode has its maximum and the ones of the LO branches starting out at 27.13 meV and 28.43 meV have a minimum in their dispersions are in the same wave vector range $0.2 \leq h \leq 0.25$. Hence, an exchange of eigenvectors, which is possible for modes of the same symmetry, between the medium energy LO modes and the LA branch cannot be excluded although the energy gap is quite large (12 - 14 meV). Note that such an exchange of eigenvectors would completely change the discussion of the bare phonon energy in sect. V. Instead of the softening of the LA branch, we would have to consider the softening of an LO phonon over more than 20 meV from the zone centre to $\mathbf{q}_{CDW}$.

For a more detailed analysis, and because both, the experimental results for phonon energies and line widths, are in good agreement with the predictions of DFPT (see sec. IV.2), we looked into the calculated phonon eigenvectors of the three investigated modes taking into account the respective electronic contributions to the phonon line width $\gamma$. The calculated absolute atomic displacements $\mathbf{u}_{atom}$ for the LO1, LO2 and the acoustic branches along the $\Gamma - M$ line are shown in Fig. 2(a)-(c). In panels (d)-(f) of the same figure, γ's (half width at half maximum, HWHM) of the three phonon branches are plotted. For the LO1 branch the maximum in γ as a function of wave vector (Fig. 2e) coincides with the maximal Nb displacement $\mathbf{u}_{Nb}$. In the acoustic branch, the strong maximum in γ at h = 0.3 – 0.35 (Fig. 2f) is accompanied by an increase of the corresponding $\mathbf{u}_{Nb}$ having, however, only a very broad peak (Fig. 2a). There is no clear correlation between the movements of the Se atoms (Fig. 2b,c) and γ values. This already indicates that the dominant part of γ is due to scattering by electronic states with Nb character, as will be shown below. The LO2 branch does not show a relation between $\mathbf{u}_{Nb}$ and γ, although there is a maximum in the former close to the zone centre (Fig. 2a).

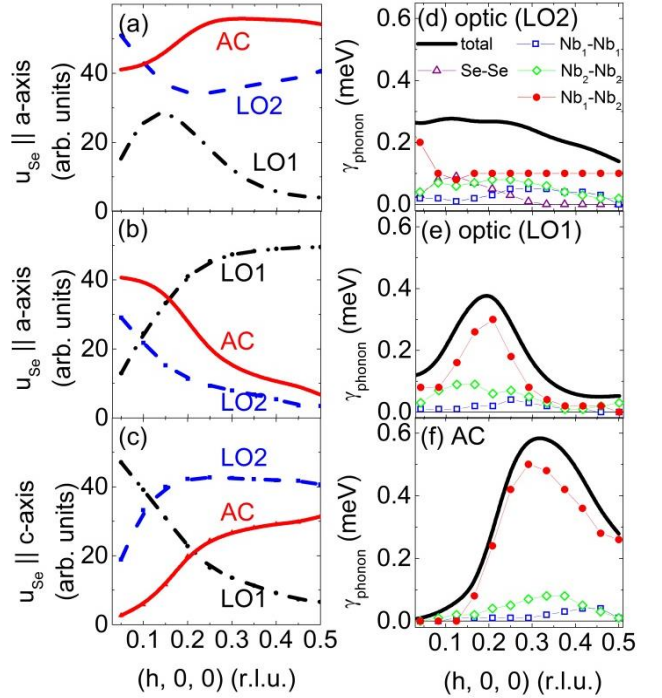

Figure 2:
*(a)-(c):* q dependence of the calculated absolute atomic displacements for the acoustic (AC), lower optic (LO1) and highest optic branch (LO2). Three displacements are shown: *(a)* Nb ∥ a, *(b)* Se ∥ a, and *(c)* Se ∥ c. Other components are zero.
*(d)-(f):* Calculated electronic contribution to the phonon line width γ for the *(d)* highest optic (LO2), *(e)* the lower optic (LO1) and *(f)* and the acoustic (AC) branches. Each panel shows the total line width (solid line) and the dominant contributions related to specific electronic scattering processes. The inset in *(e)* shows the calculated q dependence of the electronic joint density-of-states (JDOS) for the scattering process Nb$_1$ – Nb$_2$.

Here, it is instructive to look into the contributions to the total γ due to different scattering processes at the Fermi surface. Our calculated Fermi surface is produced by 2 Nb $4d$ – derived bands (which we call Nb$_1$ and Nb$_2$ for simplicity) and one band with Se $4p$ character in agreement with previous reports [23,24]. Apparently, electronic scattering paths between the two Nb *4d*-derived bands are responsible for the strong maxima in the EPC of the LO1 and acoustic modes (Fig. 2e,f). Further, we see that the same scattering path yields a weak maximum in γ for the LO2 branch close to the zone centre (Fig. 2d), which coincides with the maximum in $u_{Nb}$ for this branch (Fig. 2a). Hence, our analysis demonstrates that phonon displacements with strong Nb movements are necessary to produce a large and wave vector dependent EPC. The consecutive maxima of γ in the LO2, LO1 and acoustic branches going from close to the zone centre to the zone boundary, i.e. from h = 0 to 0.5, might indicate a certain transformation of Nb character from the LO2 to the LO1 and then to the acoustic branch along this direction. However, the wave vector dependence of the Nb displacements, in particular the one of the acoustic branch, argues against a decisive role of such an exchange with respect to the formation of CDW order. $u_{Nb}$ increases by one third at h = 0.15 – 0.35 and then is reduced by 3% further towards the zone boundary (Fig. 2a). On the other hand, the corresponding γ related to scattering between the Nb *4d*-derived bands jumps from zero at small wave vector to a clear maximum at h = 0.3 and then decreases again by 50% (red dots in Fig. 2f). We conclude that DFPT does not predict an exchange of eigenvectors between the LO and LA branches of $\Sigma_1$ symmetry in $NbSe_2$.

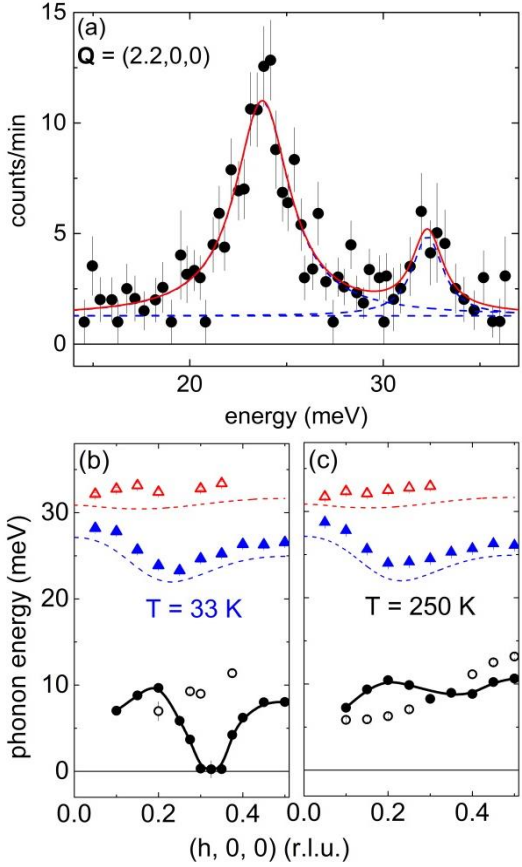

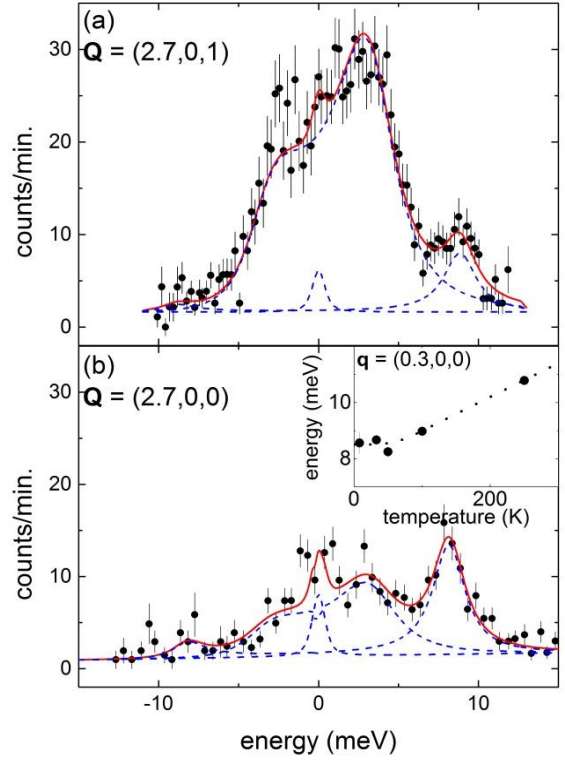

Figure 3: Raw IXS data obtained at T = 50 K and Q = (2.7,0,1) *(a)* and (2.7,0,0) *(b)*. Solid (red) lines are fits consisting of an elastic line, damped harmonic oscillators for the inelastic peaks and a linear background shown as dashed (blue) lines. *Inset:* Temperature dependent phonon energy of the second lowest longitudinal mode at q = (0.3, 0, 0) (see *(a), (b)*). The dashed line is a guide to the eye.

Figure 4: (a) Raw data showing two optic phonon modes at Q = (2.2,0,0) at T = 33 K. The solid (red) line is a fit consisting of two damped harmonic oscillators for the inelastic peaks convoluted with the experimental resolution and a linear background shown as dashed (blue) lines. (b),(c) Measured phonon dispersion at T = 33 K and 250 K, respectively. Solid lines denote the dispersion associated with the soft phonon mode (see text) as published in Ref. [Weber11]. Dashed lines are the calculated energies of the LO1 and LO2 optic branches (see Fig. 1).

### IV.2 Experimental results

Experiments using inelastic neutron scattering to measure phonons in materials having a CDW with good wave vector and energy resolution were limited to a small number of compounds, where sufficiently large single crystals could be grown [25]. Similar measurements in $NbSe_2$ [10,11] gave only limited insight into the dynamics in the pre-transitional temperature region just above $T_{CDW}$. Apart from the small sample volume, previous measurements focused on the Brillouin zone adjacent to the reciprocal lattice vector τ = (3,0,0). In contrast, our calculations predicted a much larger structure factor for the soft phonon mode around τ = (3,0,1). Fig. 3 shows IXS raw data taken in the two different Brillouin zones at a reduced wave vector of q = (0.3,0,0), which demonstrate the accuracy of the calculated structure factors and, hence, the calculated phonon pattern. Indeed, the soft mode pattern includes movements of Se along the c-axis although the CDW ordering wave vector has a zero component along the *(00l)* direction [10,19]. Here, it is instructive to know that the eigenvector of the soft mode does not quantitatively reflect the structural distortion in the CDW ordered phase. For instance, it is important to consider the superposition of the soft mode in different but equivalent directions, such as the three equivalent *(100)* direction in the hexagonal lattice of 2H-NbSe$_2$. Extracting the CDW distortion from DFPT is in principle possible but outside the scope of this paper.

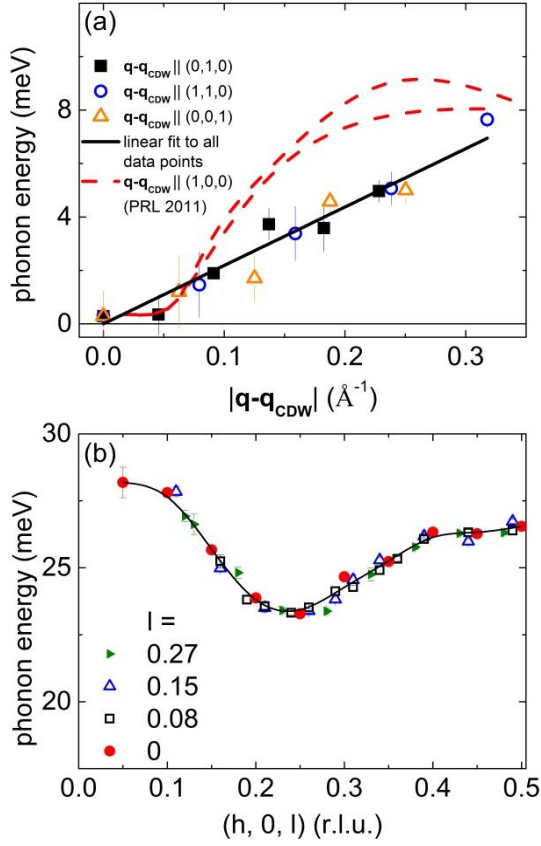

Figure 5: (a) Observed phonon energies at T = 33K along different high symmetry direction (different from [100]) starting from the CDW wave vector $q_{CDW}$. Energies are shown as function of the distance from $q_{CDW}$ in absolute units. The solid line is a linear fit of the data. The dashed lines indicate the corresponding phonon energies along the (100) direction as published in [4] (two lines for $q < q_{CDW}$ and $q > q_{CDW}$). (b) Dispersion of the LO1 branch at T = 33K for different values of l along the [001] direction. The line is a guide to the eye.

Energy scans of the optic phonon branches are shown in Fig. 4 along with the observed dispersion along the crystallographic (100) direction at T = 33 K and 250 K. We measured the optic branches corresponding to the calculated ones starting at 27.13 meV (LO1) and 30.86 meV (LO2) at $\mathbf{Q}$ = (2+h,0,0). Scans in a different Brillouin zone showed phonon peaks at slightly higher and lower energies compared to the ones of LO1 and LO2 modes, respectively, in agreement with our structure factor calculations. Due to the limited amount of beam time, however, we could not determine all dispersions completely and focused on the branches detectable at $\mathbf{Q}$ = (2+h,0,0). In addition to the dispersion of the soft phonon reported earlier [13] we include also the phonon energies of the lowest LO branch with $\Sigma_1$ symmetry. The intensity ratio of the soft mode and this second lowest energy branch is very different in the two Brillouin zones adjacent to $\boldsymbol{\tau}$ = (3,0,0) and (3,0,1) (Fig. 3) in agreement with DFPT. Therefore, an unambiguous assignment of the phonon characters of the soft mode and the lowest LO branch was possible in a simultaneous evaluation of energy scans at the same $\mathbf{q}$ value in the two different Brillouin zones, i.e. adjacent to $\boldsymbol{\tau}$ = (3,0,1) and (3,0,0). Apart from a small offset towards higher experimental energies, we see good agreement between the observed and calculated phonon energies for the LO1 and LO2 branches. In particular, the minimum in the dispersion of the LO1 mode is observed. As discussed in sect. IV.1 the concurrence of the dip in the dispersion of the LO1 branch and the maximum of the dispersion of the soft mode at h = 0.2 r.l.u. – 0.25 r.l.u. might suggest an exchange of eigenvectors between the two modes at these wave vectors. However, the dispersion of the LO1 branch does not change between T = 33 K and 250 K. This is in contrast to the huge temperature effect in the acoustic mode and speaks against a sizeable interaction of the branches in agreement with our theoretical analysis of the line width contributions.

To investigate a possible interaction of the optic and acoustic modes more closely, we measured the soft mode energies going radially away from $\mathbf{q}_{CDW}$ in various directions in reciprocal space (Fig. 5a). We found that the dispersions can be well approximated by a straight line regardless whether we move away from $\mathbf{q}_{CDW}$ within the basal plane or along the (001) direction. Results for the latter direction demonstrate that the softening occurs only in a small region of reciprocal space. In our experimental setup, consecutive analysers of the HERIX spectrometer sampled different wave vectors, which were spaced along the (h, 0, l) line in reciprocal space. Hence, we are able to construct a 2D dispersion surface of the LO1 branch in the (h, 0, l) plane. For simplicity, we show the results as line scans for different values of $0 \leq l \leq 0.27$ (Fig. 5b). Apparently, the dispersion does not depend on l. The qualitative discrepancy between the l dependence of the soft mode and the LO1 branch further indicates that an interaction between the two branches is small or absent altogether. We note that the above discussed observations are in excellent agreement with DFPT, which predicts a 4 meV higher energy of the soft mode at $\mathbf{q}$ = $\mathbf{q}_{CDW}$ + (0, 0, 0.5) (corresponds to $|\mathbf{q}-\mathbf{q}_{CDW}|$ ≈ 0.25 Å$^{-1}$ in Fig. 5a). Furthermore, the LO1 branch shows also no l dependence in DFPT.

In Fig. 6a we plot the phonon line widths along the $\Gamma - M$ direction of the optic phonons at T = 33 K and 250 K. We note that in phonon spectra taken at $\mathbf{q}$ = (h, 0, 0), h ≤ 0.1, the LO2 mode dominates, whereas for h > 0.1 the LO1 mode shows the larger spectral weight (e.g. see Fig. 4a for h = 0.2). Though the determination of the phonon energies was almost always possible for both phonons, the line widths could only be measured accurately for the dominant peaks. Therefore, we show line widths in Fig. 6a for the LO2 modes at h ≤ 0.1 and for the LO1 modes at h ≥ 0.15.

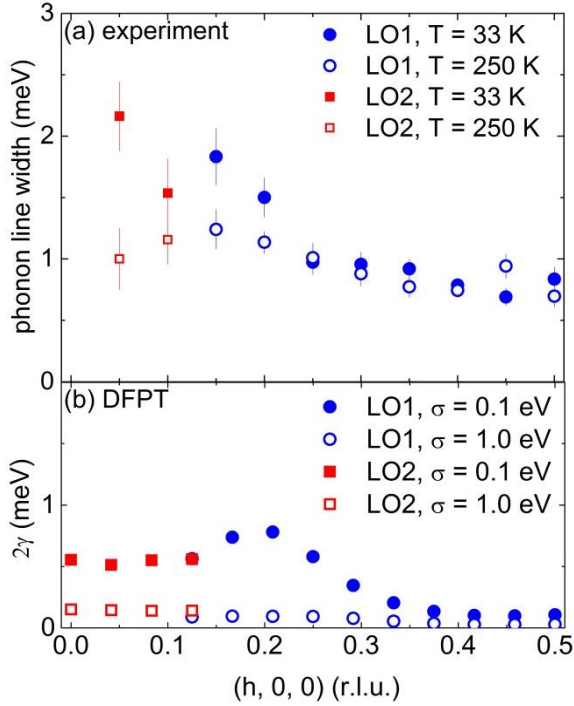

Figure 6: *(a)* Wave vector dependent line widths (FWHM) at T = 33 K (filled symbols) and 250 K (open symbols) of the LO1 and LO2 branches. *(b)* Calculated electronic contribution to the phonon line width 2γ for the LO1 and LO2 branches with two different numerical smearing parameters σ = 0.1 eV and 1.0 eV.

We observe a nearly constant line width of about 1 meV at T = 250 K, which we assign to the general imperfections of the crystal and anharmonic effects. At low temperatures, however, the line widths at wave vectors with h ≤ 0.2 increase significantly, whereas for h ≥ 0.25 we find the same values as for high temperatures.

Here, we compare our results to calculations of the electronic contribution to the phonon line width 2γ in our DFPT calculation. Although DFPT is a zero temperature technique it was shown in several publications that the necessary numerical smearing of the electronic states σ acts like a thermal smearing in electronic momentum space and, hence, can be used to qualitatively simulate temperature effects [26]. Though temperatures equivalent to σ are at least 1 order of magnitude too large, our previous investigation of the soft mode in 2H-NbSe$_2$ indicated that values of 0.1 eV ≤ σ ≤ 1.0 eV produce results that are consistent with a temperature range of 30 K ≤ T ≤ 300 K [13].

Fig. 6(b) shows the calculated values of 2γ (corresponding to the FWHM) for the LO1 and LO2 phonon modes at the wave vectors corresponding to Fig. 6(a). We see a good agreement regarding the wave vector dependence between the line widths at low temperature and 2γ calculated with σ = 0.1 eV although the decrease of 2γ is somewhat more gradual than in the observed line width. This wave vector dependence is much reduced at T = 250 K as well as in the calculations using σ = 1.0 eV. In particular, the calculations nicely reproduce the observed temperature effect at low values of h. Therefore, we assign this observation to the presence of EPC in these branches at h ≤ 0.2.

In summary, the experimental results confirm the prediction of substantial EPC in optic phonon branches at small reduced wave vectors. The results of the phonon energies at different temperatures and away from the (1, 0, 0) high symmetry line do not indicate a strong interaction between the LO1 and LA branches in agreement with DFPT.

## V. Discussion

In his seminal paper on the formation of a CDW in a one dimensional metal, Peierls pointed out that CDW order is only stabilized by a coupling of the electrons to the lattice, i.e. EPC. Hence, the investigation of the lattice degrees of freedom is a source of unique information in compounds undergoing a CDW phase transition. Theoretically, CDW materials were investigated intensively in the 1970s [7,27,28] and early 1980s [8] and, more recently, using modern *ab-initio* methods [22,29].

One important piece of information is the bare phonon energy $\omega_{bare}(q)$ in the absence of renormalization effects linked to the CDW phase transition. $\omega_{bare}(q)$ and the experimentally observable renormalized phonon frequency $\omega_q(q)$ are linked by [28]

$$\omega_q^2 = \omega_{bare}^2 - \frac{2N^3 g_q^2 Re\{\chi_q\}}{M[1+(2\bar{U}_q - \bar{V}_q)Re\{\chi_q\}]}. \qquad (2)$$

where $N, M, \bar{U}_q$ and $\bar{V}_q$ are the Avogadro constant, ionic mass, average coulomb and exchange matrix elements, respectively. $\chi_q$ is the electronic response function and $g_q$ the EPC matrix element. Typically, the last term in equation (2) is small. In CDW compounds, it is expected that $\chi_q$ becomes large at $\mathbf{q}_{CDW}$ causing the collapse of the phonon mode. However, in a recent publication [13] we have shown for the case of 2H-NbSe$_2$ that also the wave vector dependence of the EPC matrix element $g_q$ can lead to a CDW instability.

For a better understanding it is instructive to explain in more detail what is the bare phonon energy $\omega_{bare}$. The originally very high phonon energies of the lattice of ionic nuclei are reduced to the typical values in the meV range by the general screening of all electrons in a solid. Our interest lies in the additional screening due to the presence of the CDW phase transition. These processes involve primarily the bands forming the Fermi surface. Therefore, we consider in our analysis only the contributions of the three bands crossing $E_F$ to the electronic susceptibility $\chi_{q,FS}$, the real part of which was calculated, e.g., by Johannes et al. [24]. Accordingly, $\omega_{bare}$ can be related to the observed energy $\omega_q$ by equation (2) if we replace $Re\chi_q$ by $Re\chi_{q,FS}$. The **q** dependence of the difference between $\omega_q^2$ and $\omega_{bare}^2$ is governed by $Re\chi_{q,FS}$ and/or $g_q$, the EPC matrix element involving the respective phonons and electronic scattering processes. In a previous publication [13] we argued that, in the case of 2H-NbSe$_2$, it is the latter, which determines the soft mode dispersion and defines the periodicity of the ordered phase, i.e. $\mathbf{q}_{CDW}$. This point of view is corroborated by the new results presented in this publication.

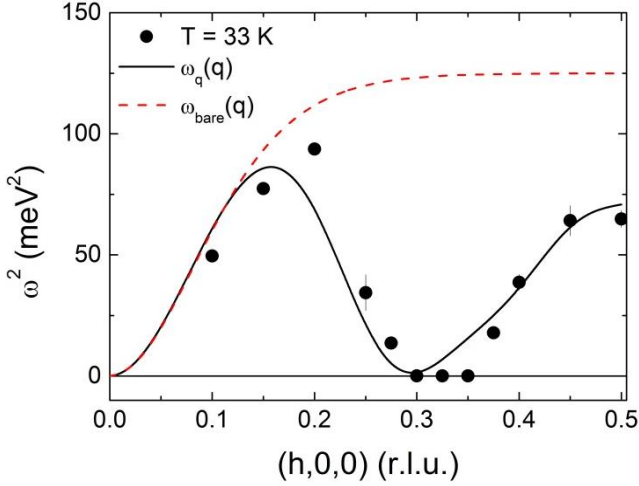

Figure 7:
(a) Fit (solid line) of the observed square soft mode energies at T = $T_{CDW}$ (dots) using equation 3 and $Re\,\chi_{q,FS}(h)$ and $g_q(h)$ as extracted from DFPT. $\omega_{bare}$ (dashed line) was approximated by a Brillouin function with two fit parameters.

In our results section we showed that an exchange of eigenvectors between LO and LA phonons with $\Sigma_1$ symmetry is very unlikely and, hence, an acoustic-like assumption for the bare phonon dispersion $\omega_{bare}(q)$ is justified for NbSe$_2$.

Before we can apply equation (2) in order to estimate $\omega_{bare}(q)$ we need to take into account the fact that the $\mathbf{q}$ dependence of $\gamma$ is largely due to scattering between the two Nb bands at the Fermi surface and that the strong wave vector dependence of $\gamma$ can originate in the corresponding EPC matrix element and/or electronic JDOS [30]. The latter varies only by ±15% along $\Gamma - M$ (not shown). Thus, the matrix elements for the Nb$_1$ – Nb$_2$ inter-band scattering indeed exhibit the strong wave vector dependence. This is in good agreement with experiments using angle-resolved photoemission spectroscopy (ARPES), which reported the highest EPC strength on the double-walled Nb 4d-derived FS sheet [31].

Therefore, we can extract in a first approximation the wave vector dependence of the EPC matrix element from our measurements and apply this knowledge in equation (2). The approximation originates in the fact that $\gamma$ is actually proportional to the square of the momentum averaged EPC matrix element $|\overline{g_q}|^2$ [30],

$$\gamma_q \propto |\overline{g_q}|^2 \times JDOS. \qquad (3)$$

However, $\overline{g_q}$ should in general have a weaker wave vector dependence than $g_q$, i.e. we expect to underestimate the real wave vector dependence of $g_q$ if at all. Concerning our discussion of which electrons should be taken into account in order to analyse the renormalization at the CDW phase transition, we will use the wave vector dependence of $Re\,\chi_{q,FS}$ as calculated by Johannes et al. [24], where only contributions from the three electronic bands crossing the Fermi energy were taken into account. It is also in good agreement with estimates based on ARPES experiments [32].

Fig. 7 displays our analysis of the renormalized and bare phonon dispersions based on equation (2) in comparison to the experimental energies at T = 33 K and 250 K. We use equation (2) in a parameterized form,

$$\omega_q^2 = \omega_{bare}^2(h) - \frac{c_1 \times f_g(h)}{\left(\frac{1}{Re\,\chi_{q,FS}(h)} + c_2\right)} \qquad (4)$$

where $\omega_{bare}$ was approximated by a Brillouin function with two parameters[1]. $f_g(h)$ and $Re\,\chi_{q,FS}(h)$ are the wave vector dependences of $g_q^2$ and $Re\,\chi_q$, respectively. The latter was taken from Ref. [24] (Fig. 7). For $f_g(h)$ we used the functional form of the contribution to $\gamma$ calculated for the acoustic mode, which is due to scattering between the two Nb 4d-derived bands at the Fermi surface (red dots in Fig. 2f). The fit was constraint by $c_2 = 2\overline{U}_q - \overline{V}_q$ having a positive value. It turned out to be a very small number the exact absolute value having negligible influence on the resulting bare dispersion in agreement with assumptions in Ref. [28]. The parameter $c_1$ takes care of the prefactors given in equation (2) and the fact that the wave vector dependent function $f_g(h)$ is only proportional to the EPC matrix element $g_q^2$ (see above).

Results of our analysis shown in Fig. 7 demonstrate that the renormalized phonon dispersion at T = $T_{CDW}$ can be well described based on an acoustic bare phonon dispersion with a zone boundary energy of 11.2 meV. The fitted bare dispersion is in reasonable agreement with phonon energies measured near room temperature although the dispersion at T = 250K is still renormalized, which is evident in the clearly observable dip around $\mathbf{q}_{CDW}$ [13]. The fact that our analysis of the phonons in NbSe$_2$ based on lattice dynamical calculations using DFPT yields a sensible bare phonon dispersion is further evidence for the accuracy of the model and the already in Ref. [13] proposed point of view that the wave vector dependence of the electron-phonon coupling drives the CDW transition in this compound. Very recently this was also corroborated by results from scanning tunneling microscopy (STM) [33]. Apart from the usually observed triple-$\mathbf{q}$ CDW wave vector $\mathbf{q}_{CDW}$ = (0.329, 0, 0), STM revealed surface regions with a unidirectional CDW wave vector $\mathbf{q}_{1Q} \approx$ (2/7, 0, 0). We note that the minimum of the fitted soft mode dispersion at T = $T_{CDW}$ is close to the value of h = 2/7. However, we do not see a clear link between our results and the single-$\mathbf{q}$ CDW order observed by STM [33]. Experimentally, inelastic x-ray scattering is a bulk probe and the DFPT calculations even overestimate the position of the CDW wave vector along the $\Gamma - M$ direction with the dispersion minimum at **q** = (0.375, 0, 0) (Fig. 1a, see also Fig. 4 in Ref. [13] for more details).

---

[1] It can be easily shown that a Brillouin function gives a better description of a typical acoustic dispersion than the simpler sine function.

## VI. Conclusion

We have reported an inelastic x-ray scattering and ab-initio theoretical investigation of the lattice dynamics of NbSe$_2$ focusing on longitudinal phonon modes across the CDW ordering wave vector **q**$_{CDW}$ = (0.329,0,0), i.e. in the (100) direction. We derive an acoustic-like bare dispersion $\omega_{bare}(q)$ of the CDW soft phonon mode from the wave vector dependence of the EPC matrix element $g_q$ and the response of the electrons forming the Fermi surface. Although, our measurements provide evidence for EPC in optic branches as well, we demonstrate that there is no significant inter-mode hybridization. Further, our analysis shows that the observed wave vector dependent EPC originates from one particular electronic scattering process between two Nb – *4d* derived bands at the Fermi surface. Together with a corresponding quasi-constant electronic JDOS for these scattering processes, this is evidence of the strong wave vector dependence of the EPC matrix elements.

## Acknowledgements

We acknowledge valuable discussions with J. van Wezel. We thank John M. Tranquada for supplying us with a single crystal of 2H-NbSe$_2$. F. W. was supported by the young investigator group "Competing phases in superconducting materials" of the Helmholtz Society (VH-NG-840). D.R. was supported by the DOE, Office of Basic Energy Sciences, Office of Science, under Contract No. DE-SC0006939. Work at Argonne was supported by U.S. Department of Energy, Office of Science, Office of Basic Energy Sciences, under contract No. DE-AC02-06CH11357.